# Exploring the acceptability of digital contact tracing for UK students

DAVE MURRAY-RUST, TU Delft
LUIS SOARES and KATYA GORKOVENKO, University of Edinburgh
JOHN ROOKSBY, Northumbria University

Contact tracing systems control the spread of disease by discovering the set of people an infectious individual has come into contact with. Students are often mobile and sociable and therefore can contribute to the spread of disease. Controls on the movement of students studying in the UK were put in place during the Covid-19 pandemic, and some restrictions may be necessary over several years. App based digital contact tracing may help ease restrictions by enabling students to make informed decisions and take precautions. However, designing for the end user acceptability of these apps remains under-explored. This study with 22 students from UK Universities (inc. 11 international students) uses a fictional user interface to prompt in-depth interviews on the acceptability of contact tracing tools. We explore intended uptake, usage and compliance with contact tracing apps, finding students are positive, although concerned about privacy, security, and burden of participating.

CCS Concepts: • **Human-centered computing** → *Empirical studies in HCI*; **Smartphones**; • **Social and professional topics** → **Governmental surveillance**; • **Security and privacy** → **Social aspects of security and privacy**.

Additional Key Words and Phrases: digital contract tracing, acceptability, design implications, interview study

## 1 INTRODUCTION

Contact tracing is "the process of identifying, assessing, and managing people who have been exposed to a disease to prevent onward transmission" [2]. Contact tracing has been used to contain sexually transmitted diseases [15, 21], for dealing with outbreaks of airborne infections in humans including SARS [16], and continues to "play an essential role in reducing the number of infections and saving lives" during the Covid-19 pandemic [36].

Traditionally, contact tracing has involved asking infected people to recall relevant contacts they have had, and then notifying those contacts that they may be infected and should take appropriate precautions such as being tested. This is resource intensive, can be slow, is prone to bias, and is limited to contacts that the infected person knows and has details for (for example, a person is unlikely to have the details of a stranger they have sat next to on public transport). More recently, methods of digital contact tracing have been developed. Digital contact tracing draws upon pervasive computing infrastructure such as mobile devices and Bluetooth to automatically detect when people have contact, to store details of this, and to later alert them if they are potentially infected.

For any public health intervention to be successful, it must be *acceptable* to those it is intended to reach. A recent UK report has underlined the importance of acceptability for the uptake and success of digital contact tracing:

> "The effectiveness of a digital contact tracing app will be contingent on widespread public trust and confidence, which must translate into broad adoption of the app." [3].

Contact tracing therefore is not just a technical problem of collecting data and notifying people, but a sociotechnical one in which design and human values must align. There is therefore a need for research to understand personal attitudes to contact tracing solutions [19].

Authors' addresses: Dave Murray-Rust, d.s.murray-rust@tudelft.nl, TU Delft; Luis Soares; Katya Gorkovenko, University of Edinburgh; John Rooksby, Northumbria University.

In this paper we investigate the acceptability of digital contact tracing solutions to one important demographic: University students. We use design methods and in-depth interviews with 22 students to explore their perspectives and attitudes towards digital contact tracing in the context of the Covid-19 pandemic. All interviewees are or have recently been studying at Universities in the UK. Half are UK or EU national students, and the other half are 'international students' who ordinarily live in mainland China. Working with this cohort is important because: students are often highly mobile and social and therefore potential spreaders of viruses; students and young people in the UK are currently 'low priority' for vaccination; and because international students account for a large proportion of people studying in the UK (almost half a million in 2018/19, of which 120,385 were nationals of China [45]).

Broad adoption of appropriately designed digital contact tracing solutions among students may be one important aspect of 'returning to normal' in university life. This process may take several years even after mass vaccination. Contact tracing apps may also become increasingly important in managing other ongoing diseases and future pandemics. Therefore, the paper unpacks four general themes with respect to the acceptability of digital contact tracing to university students:

- What factors affect students' support for the use of contact tracing apps? This issue is important to explore because understanding the motivations for participating in an intervention can help shape the design and messaging surrounding it. This work can therefore influence how contact tracing is presented and what concerns should be addressed.
- What concerns do students have about contact tracing apps? While there has been public debate on various aspects of contract tracing, particularly privacy, it is important to include end-user perspectives in the discussion. This work therefore uses in-depth interviews to unpack and understand students real-world concerns
- What might affect students' decisions to comply with instructions from a contact tracing app to self isolate? Current contact tracing apps serve two functions: to trace the spread of disease, and to let users know they may have had a contact and should self isolate. This is burdensome, so it is important to understand if and how users would follow difficult advice given.
- Why might students uninstall a contact tracing app? This issue is important because even if students install an app they may choose to remove or disable if it is not acceptable to them.

To organise the findings and in order to gain a comprehensive picture of the responses of participants, this paper will use the Theoretical Framework for Acceptability (TFA) [40] to structure the results. The TFA is used to direct analysis towards multiple factors including; the "affective attitude" held towards the intervention, and the perceptions of "burden", "opportunity cost", "efficacy" and "coherence" of digital contact tracing.

## 2 BACKGROUND

Covid-19 (Coronavirus disease 2019) is a new and highly infectious disease that first emerged in China in December 2019, and has was declared a global pandemic by the WHO in March 2020. On the 13 March 2020, the WHO announced that they considered Europe to have become the active centre of the pandemic. An initial lockdown in the UK was announced on the 23rd March 2020, with the Prime Minister "telling residents they must leave their home only to travel to work where 'absolutely necessary,' to shop for essential items, to exercise once a day, and to access medical care. Shops selling non-essential goods were told to close immediately" [25]. The UK has since been through several lockdowns.

## 2.1 Covid-19 and university students

For students studying in the UK, the global pandemic and associated lockdowns meant that, in the first half of 2020, many courses were cancelled or moved online. Many students decided to leave their accommodation early to return 'home', although others remained through choice and some were stranded [48]. Some UK Universities were criticised for "provoking wide-scale panic among international students" [11] through an incoherent approach to lockdown and advice to leave. In the second half of 2020, many university courses were taught online and/or with social distancing measures. This is causing uncertainty for students.

Many students are young adults and thus not at high direct risk from Covid-19. However, students in the UK and around the world have been experiencing distress. A study in Italy found that lockdown had a greater negative impact on student's sleep quality than for university staff, and that there were increases in anxiety [31]. Studies from Spain and Greece have suggested students were at high risk of mental distress during lockdown [35, 38]. A report from Australia has highlighted the financial burdens and precarity of international students [13], and found visa concerns meant international students were reluctant to seek help. In Australia a quarter of international students had experienced racism during the pandemic, with a substantially higher proportion of Chinese students receiving abuse. Similar problems are evident in the UK [47], although there were greater protections for migrants in the UK than several other countries [13].

## 2.2 Digital contact tracing

Contact tracing is one of the public health measures for Covid-19 recommended by WHO. Contact tracing is a well established method of managing infections in human and non-human populations [15, 18, 21]. There is a rapidly growing body of literature about the application of contact tracing to the Covid-19 pandemic, with some key summaries, in particular: the World Health Organisation's general report [2] and ethical considerations [49]; the Ada Lovelace Institute's Rapid Evidence Report [3]. Digital contact tracing potentially replaces large numbers of human contact tracers carrying out phone interviews and diary recall exercises with pervasive technology that does the same job at scale with less human intervention [42].

*2.2.1 Digital contract tracing apps in the UK and China.* The response to the pandemic across the UK has been organised and coordinated by the governments and health services in Scotland, Northern Ireland, England and Wales. Three national digital contract tracing apps were developed:

- NHS Covid-19: This app was developed for use in England and Wales. An initial version trialed in May 2020 collected data to a centralised server, but this was abandoned [46]. A decentralised version developed by a private company on behalf of the NHS was released in Sept 2020 [17]. This version tracks contacts, alerts when to self isolate and provides a timer, gives information about virus in local area, and allows 'check ins' to venues such as pubs and restaurants.
- Protect Scotland: This app was developed for the Scottish Government and released for use in Scotland in September 2020 [22]. The app has a simpler design than NHS Covid-19, providing alerts when a user has been in contact with someone diagnosed with the virus.
- StopCOVID NI: This app was developed for Health and Social Care in Northern Ireland. This app served as the basis for Protect Scotland and has strong similarities.

Analysis of the Covid-19 app [14] indicates that it has been downloaded 21.63 million times (representing 56% of over 16s with smartphones). by January 2021 notifications to isolate had been sent to 263,000 app users in England and 8,600 in Wales [14].

None of the UK apps are mandatory and their use has sometimes been controversial. Several employers across the UK asked workers not to use the apps because of concerns about false positive rates. UK Universities encourage use of the apps but also make use of other campus apps for monitoring building use and monitoring attendance. Ultimately, most students across the UK have been asked not to attend University in person for a large proportion of time since the pandemic began and therefore the potential for contact tracing in this context remains largely unverified.

Because this study includes international students with family links to China and who in many cases returned to China during lockdowns, it is appropriate to discuss technology used in China. A system known as "Health Code" has been used since May 2020 that builds upon existing platforms AliPay and WeChat. This provides a QR code and a colour code serving as a 'pass' that can be scanned by authorities and when accessing buildings and areas. This is not specifically a contact tracing app, but is used to ask health information, track movement, and designate levels of risk. It uses a large amount of data, including GPS logs, although the exact mechanisms are not public [43].

*2.2.2 Effectiveness of contact tracing.* A Cochrane Rapid Review reminds us that there is still little evidence regarding the effectiveness of contact tracing [9]. Several modelling studies (e.g. [20, 28, 30]) and some practical experience [42] indicate that digital contract tracing could be a useful intervention in the context of the Covid-19 pandemic. However, computational modelling disagrees about the effectiveness of digital contact tracing, with some studies indicating that 80% of smartphone users using an app would suppress the pandemic [24], or that it can be effective in slowing down the pandemic [29], while others find that it is unlikely to have an effect. It should be said that the effectiveness of traditional forms of contact tracing has also been questioned for different diseases and populations [27] [41].

*2.2.3 Ethicality of contact tracing.* Contact tracing can be ethically and socially problematic due to a number of issues, including privacy associated with population-level surveillance [12, 23]. Ethical concerns have also been raised [7, 26, 34] including that digital contact tracing will more readily benefit people in more affluent areas with people who can afford mobile devices capable of running the technology.

*2.2.4 Technical issues in contact tracing.* Digital contact tracing poses several technical challenges to do with continuous data collection and matching, against a background that "surveillance can quickly traverse the blurred line between disease surveillance and population surveillance." [49, p.1]. A key debate in the technology community has been between solutions that centralise data processing, and decentralised approaches such as DP3-T [5, 44] that maintain strong guarantees on privacy by minimising data sharing. In response to this, Apple and Google cooperatively developed operating system level support for decentralised contact tracing that some apps, including the NHS Covid-19 and Protect Scotland apps build upon [10]. Several countries do not use this, for example Australia uses a decentralised "Bluesafe" protocol. Beyond privacy, there are concerns about technical functionality such as the reliability of Bluetooth signal strength as an indicator for viral transmission – although research indicates more robust measures are possible [1]. Technical and security problems have also hit several contact tracing apps. For example the initial release of the Covid-19 app in the UK did not get past a trial stage [46], and an app in Japan was not able to accept codes from infected users.

## 2.3 Acceptability of contact tracing

Since the envisaged effectiveness of contact tracing is dependent on population scale uptake, the acceptability of the intervention to end users is of high importance. The largest public opinion survey so far published on the acceptability of contact tracing apps concluded that the "main reasons against installation are (i) an increased risk of government surveillance after the epidemic, (ii) that one's anxiety about the epidemic would increase, and (iii) fear of one's phone being hacked" [33]. Another has found "strong support" for digital contact tracing [8]. This gives a sense of how large numbers of people may act, but there are two significant areas for interrogation. Firstly, this captures stated intent, rather than actual behaviour, which is of particular concern when some potential actions such as ignoring a request to self isolate may be illegal. Secondly, they do not capture the range of design decisions and features present in a real application. Cultural and political differences may also affect the uptake of digital contact tracing– for example, in the words of one commentator, there are "a number of features of China's response to COVID-19 that would be unlikely to be effective or acceptable in other countries." [37].

Beyond questionnaires, acceptability in digital health is typically invested qualitatively to understand positive and negative aspects of existing interventions [32]. An emerging tool for looking at acceptabilitly is Sekhon et. al's Theoretical Framework for Acceptability (TFA)[40]. This framework is designed to make sure that all of the facets of an intervention are taken into account - not simply a users views on its ethics, but the practical concerns of how difficult it is to participate, how well they perceive it functioning and so on. This helps to build up a broad picture how people relate to technology, with previous work [39] showing that even for an ethcially charged intervention such as predicting mental health from smartphone behaviour, mundane concerns such as battery life were a strong factor in decision making [39]. A study on app based contact tracing using focus groups, before the widespread release of the apps was able to identify key factors for uptake around privacy, stigma, misconceptions and altruism [50].

## 3 THE STUDY

For this study we have carried out in-depth qualitative research with 22 participants. We used mock interfaces of digital contact tracing apps as prompts.

### 3.1 Materials

A mock interface for a contact tracing app was developed collaboratively by all four authors for use as a study prompt. The interface designs were created to support consideration and discussion rather than appear to be a single cohesive application. The mock interface was designed to include elements that could help illuminate participants' attitudes and opinions, including to more provocative and controversial issues. It is not the aim of this study to assess the acceptability of one existing application (e.g. one of the three UK apps) and our designs hold both similarities as well as differences to those.

Our design work sought to include novel and provocative features that we believe are conceivable for contact tracing, including: offering behavioural advice, displaying local or hyper-local information and offering various kinds of encouragement or information. In our design work we also sought to account for a set of concerns derived from the design space outlined in a review [3, p.22]: what data is collected, who can access the data, how contacts are alerted, and what actions the user is asked to take. Our design work was also specifically oriented to the concerns in the Theoretical Framework for Acceptability [40],

The mock interface prompts were organised on a design board with three sections themed upon:

- The **Installation Process** (Figure 1) asked the user for various permissions as a way to foreground questions around data, privacy and usage factors: what data was collected, who could process it, and the impacts of running the app such as battery life and control of Bluetooth.
- Daily or weekly **Notifications** that the app might present to shape behaviour and maintain active participation. This included simple encouragement and national information (Figure 2) but extended to more speculative behavioural suggestions and hyper-local information (Figure 3)
- **Contact notifications** started with blunt commands to self isolate, and added in supporting information, personal risk and spatio-temporal contact logs (Figure 4).

We also created background information for participants describing: i) contact tracing and its goal; ii) data sharing; iii) centralised and decentralised digital contact tracing protocols.

### 3.2 Methods

Each participant completed the following two components of the study:

- A short questionnaire covering general background demographics, overview of living situation and any changes made in response to the pandemic, as well as responses on a 5 point Likert scale to the "Fear of COVID-19 Scale" [6], which collects standardised responses about their psychological relationship to the ongoing pandemic.
- A video interview, lasting approximately one hour, that covered background information and then used a mock interface as a prompt for a semi-structured interview about the participants attitudes to various aspects of contact tracing apps.

In the interviews the mock screens were presented in sequence using Figma software (figma.com). Discussion was structured around each screen, with an open discussion held after each section. After the final screen, a final discussion covered wider issues, ideas and concerns the student may have.

The interviews were then transcribed and coded in full. In phase one of the analysis we coded in line with the features of the mock interface. This enabled us to discover what features and functionalities the students valued. In phase two we coded inductively to draw out emergent and cross-cutting themes.

Ethical approval was secured through the [ANON] institutional review board (number xxxx-xxxx-xxxx).

### 3.3 Recruitment

We advertised for participants on university mailing lists, and then carried out snowball sampling. Participants were informed that the study was voluntary and were compensated (£25 voucher or equivalent).

We chose to sample 11 'home' students from the UK/EU and 11 'international' students from China in order to gain diverse opinion. The study is not intended to be a comparison between UK and Chinese perspectives but values and takes seriously global experiences and perspectives on student life. The majority of participants had studied in Scotland, but several were or had recently been students in England.

## 4 FINDINGS

The findings are based on interviews with 22 participants, 11 from the UK (identified by P1-P11) and 11 from China (P11-P22), with 16 identifying as female and 5 as male. They were all studying or had recently completed degrees at various levels, 7 Bachelors, 11 Masters and 4 PhDs. Most respondents lived with 3 or 4 other people (15) and one lived alone. Only one was 18-20 years old, with most (15) being 20-25, and 6 being 25-30. UK students mostly lived with

Exploring the acceptability of digital contact tracing for UK students 7

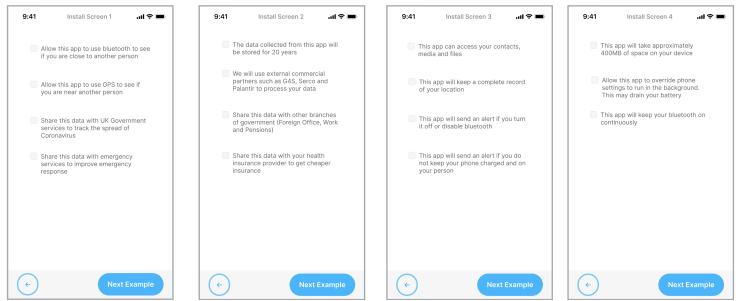

Fig. 1. The first four screens shown to the participants echo the permissions that users might be asked to give, covering a) direct questions about data collection and use; b) data retention and sharing; c) use of other personal data and control of phone features; d) possible impacts the app might have on the user

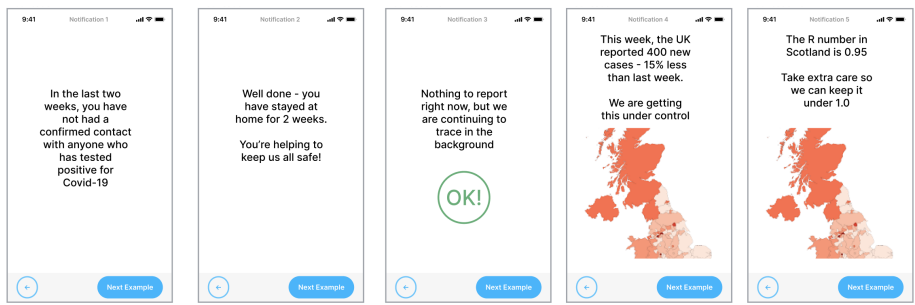

Fig. 2. Initial passive notification screens shown to users, showing a) simple statement of no contact b) encouragement for good behaviour; c) reminder that the system is functioning; d) national statistics and encouragement; e) national statistics using the $R$ number.

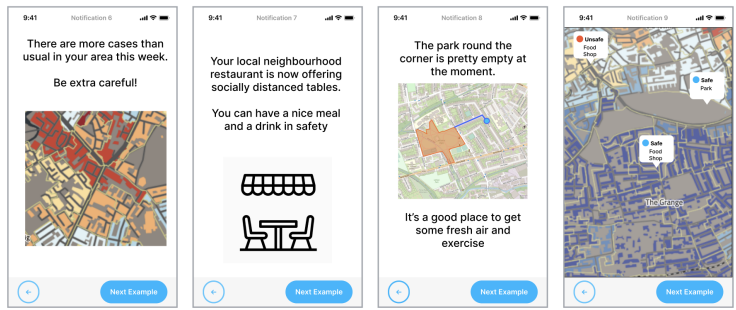

Fig. 3. A further passive notification screens shown to users, showing a) a locally targeted warning of increasing risk; b) a behavioural advert; c) a specific suggestion of place and healthy activity; d) precise classifications of particular locations as more or less risky.

parents (5), siblings (4) or flatmates (4). Chinese students were more likely to live with parents (7), but also grandparents (2), siblings (1), flatmates (2) and alone. Table 1 shows the actions that respondents had taken action in response to Covid-19; 17 had changed their living situation, while 12 had moved home, and most people (14) had self isolated or knew someone who had (16). 10 of the Chinese respondents had been in official quarantine, compared to 2 from the UK, while 9 had self isolated compared to 5 from the UK . Against this background, the Fear of Covid scale [6] indicated an medium level of concern (3.51 on a scale from 1 to 5), broadly similar across the population, without extreme outliers.

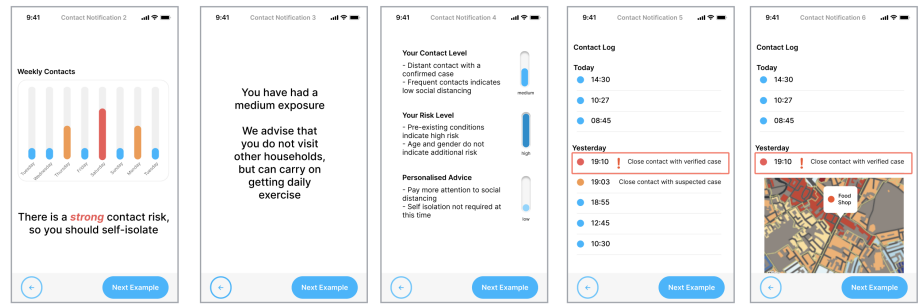

Fig. 4. A range of contact notification screens, with a) weekly summary; b) statement + instruction; c) personalised risk; d) temporal log; e) temporal log with map.

Table 1. Table of participants, showing demographic information, household size and the length of interview. In addition, the table shows actions taken in response to Covid-19 - all participants have had to make some modification to their lives, and many have had to self isolate. The table aslo shows scores on the Fear of Covid scale from 1 to 5 for: "I am most afraid of Covid-19", "It makes me uncomfortable to think about coronavirus-19", "My hands become clammy when I think about coronavirus-19", "I am afraid of losing my life because of coronavirus-19", "When watching news and stories about coronavirus-19 on social media, I become nervous or anxious", "I cannot sleep because I'm worrying about getting coronavirus-19", "My heart races or palpitates when I think about getting coronavirus-19".

| | | | | | | | Actions Taken | | | | | Fear of Covid | | | | | | | |
|---|---|---|---|---|---|---|---|---|---|---|---|---|---|---|---|---|---|---|---|
| ID | Location | Age | Gender | Study Level | Interview Length | HH Size | Changed Living Situation | Moved Home | Self Isolated | Quarantined | Acq. Self Isolated | Afraid | Uncomf. | Clammy | Losing | News | Sleep | Heart | Overall |
| P1 | UK | 25-30 | f | PhD | 0h53 | 2 | | | ✓ | | | 4 | 5 | 2 | 4 | 4 | 2 | 2 | 4.6 |
| P2 | UK | 25-30 | m | PhD | 0h59 | 6 | ✓ | ✓ | ✓ | | ✓ | 3 | 2 | 1 | 2 | 3 | 1 | 1 | 2.6 |
| P3 | UK | 25-30 | f | PhD | 1h02 | 6 | | | ✓ | | | 2 | 4 | 2 | 1 | 3 | 1 | 1 | 2.8 |
| P4 | UK | 20-25 | f | UG | 0h49 | 4 | ✓ | | | | ✓ | 2 | 4 | 1 | 1 | 3 | 2 | 1 | 2.8 |
| P5 | UK | 20-25 | m | UG | 0h56 | 4 | ✓ | | | | | 4 | 4 | 2 | 2 | 4 | 2 | 2 | 4 |
| P6 | UK | 20-25 | f | M | 1h08 | 3 | ✓ | ✓ | ✓ | ✓ | ✓ | 2 | 2 | 2 | 1 | 3 | 2 | 2 | 2.8 |
| P7 | UK | 25-30 | m | PhD | 1h33 | 2 | | | ✓ | ✓ | ✓ | 2 | 1 | 1 | 1 | 2 | 1 | 1 | 1.8 |
| P8 | UK | 20-25 | f | UG | 0h56 | 4 | ✓ | ✓ | | | ✓ | 4 | 5 | 3 | 2 | 4 | 1 | 1 | 4 |
| P9 | UK | 20-25 | f | UG | 1h09 | 4 | ✓ | ✓ | | | ✓ | 2 | 4 | 3 | 2 | 5 | 1 | 2 | 3.8 |
| P10 | UK | 20-25 | f | UG | 1h07 | 3 | ✓ | | | | ✓ | 2 | 4 | 1 | 2 | 5 | 1 | 2 | 3.4 |
| P11 | UK | 20-25 | m | UG | 0h53 | 3 | ✓ | | | | | 2 | 4 | 2 | 2 | 4 | 1 | 1 | 3.2 |
| P12 | CN | 20-25 | f | M | 1h13 | 3 | ✓ | ✓ | ✓ | ✓ | ✓ | 3 | 5 | 1 | 1 | 4 | 1 | 1 | 3.2 |
| P13 | CN | 20-25 | f | M | 1h03 | 2 | ✓ | ✓ | | ✓ | | 4 | 2 | 2 | 4 | 2 | 1 | 1 | 3.2 |
| P14 | CN | 20-25 | f | M | 1h32 | 4 | ✓ | ✓ | ✓ | ✓ | ✓ | 3 | 2 | 1 | 3 | 3 | 2 | 3 | 3.4 |
| P15 | CN | 20-25 | f | M | 1h14 | 3 | ✓ | ✓ | ✓ | | ✓ | 2 | 3 | 2 | 2 | 4 | 3 | 2 | 3.6 |
| P16 | CN | 20-25 | f | M | 1h07 | 3 | ✓ | ✓ | ✓ | ✓ | ✓ | 5 | 3 | 2 | 5 | 4 | 2 | 2 | 4.6 |
| P17 | CN | 20-25 | f | M | 1h05 | 4 | ✓ | ✓ | ✓ | ✓ | ✓ | 3 | 2 | 1 | 2 | 3 | 1 | 1 | 2.6 |
| P18 | CN | 20-25 | f | M | 1h06 | 4 | | ✓ | | ✓ | | 2 | 3 | 1 | 3 | 3 | 1 | 1 | 2.8 |
| P19 | CN | 20-25 | f | M | 1h05 | 3 | ✓ | | ✓ | ✓ | ✓ | 4 | 5 | 2 | 2 | 4 | 2 | 1 | 4 |
| P20 | CN | 25-30 | m | M | 1h10 | 1 | ✓ | ✓ | ✓ | ✓ | ✓ | 4 | 4 | 2 | 3 | 4 | 4 | 4 | 5 |
| P21 | CN | 18-20 | f | UG | 1h19 | 6 | ✓ | | ✓ | ✓ | ✓ | 3 | 4 | 2 | 3 | 4 | 2 | 2 | 4 |
| P22 | CN | 25-30 | f | M | 1h05 | 3 | | | ✓ | ✓ | ✓ | 4 | 4 | 3 | 2 | 4 | 4 | 4 | 5 |

In the rest of the findings, we report participants' experiences through the pandemic, followed by responses to the prototype, in particular support for and concerns around digital contact tracing, willingness to follow guidance and reasons for uninstalling or removing the application.

### 4.1 Feedback on the mock interface screens

Table 2. Main findings from installation screens (Figure 1) and conversations about data protection.

| Function | Main findings | Key quotes |
|---|---|---|
| Data collection | Basic data collection uncontroversial but GPS a worry. Reluctant to share full movement records or contacts Asking for extra information caused distrust | *"I don't ever have my Bluetooth turned on, so I don't think [Bluetooth] would be particularly useful for me"* [P9] *"[I started to distrust the person who's made the app slightly ... It made me question actually every single setting"* [P7] |
| Data Sharing | Worries about commercial processing and government overreach. Retention time a worry for some but not others Cheap heath insurance divisive | *"I've seen already the start of the NHS starting to get sold off"* [P5] *"if I knew the data was being stored safely, then I wouldn't have any issue of the length of time it was stored for"* [P11] *"I'm not sure whether the external commercial partners will protect my data, because in China, many companies will share [...] this data to other fake companies"* [P15] |
| Phone operation | App size not a big worry, but battery life a clear issue Disliked loss of control | *"I don't think that the app itself should have control over whether Bluetooth is on or not because there might be other reasons that you don't want it on."* [P10] |

After presenting the initial information about contact tracing, participants were positive about the idea of participating, with a sense that *"it makes you feel like you're doing something good for the country"* [P5] , or that the app *"can be helpful and give much clear information"* [P11] . There was broad understanding of the purpose: *"I don't understand all the tech behind it but I think explaining that it works by Bluetooth and measuring the distance between the phones, sort of, clears that up"* [P4] . Occasional misconceptions among a few of the Chinese participants that contact tracing was a way to *"find whether the people around you have Covid-19 and maybe to see their trace"* [P13] were cleared up as discussion progressed.

Our first phase of coding and analysis followed the structure of the mock interfaces we presented. The main findings and key quotes from this phase are presented in three summary tables. The tables cover (i) responses to the data collection, data sharing and phone operation sections of the installation screen (Table 2); (ii) the daily notifications, with context, location specific information, advertising and behavioural suggestions (Table 3), and (iii) contact alerts with various levels of detail and personalisation (Table 4).

Our second phase of coding and analysis developed inductive themes emerging from the material. These were organised and prioritised with relation to key questions: factors that support or undermine participation in digital contact tracing, factors around compliance with advice or commands, and triggers for participants to uninstall the app and end their participation in the scheme. The following Findings subsections discuss these inductive themes.

Table 3. Main findings from showing potential daily or weekly background notifications (Figure 2, 3).

| Function | Main findings | Key quotes |
| --- | --- | --- |
| Notifications in general | Valued reassurance, although over-positive messages were suspect<br>Motivation to continue<br>Even positive messages could cause anxiety | *"On one hand it shows that the app, like you are constantly tracing which is a good thing, but on the other hand it's like that would make me want to keep going back and checking my phone every five minutes."* [P11] |
| Contextual Notifications | Participants asked more questions when presented with information.<br>Fine grained information supports behaviour change<br>Summary statistics (e.g. R number) can dominate reactions, and are not always understood | *"Firstly, it is just telling me it is 15 per cent less than last week but it does not tell me… getting this under control, report 400 new cases. There is a problem that whether we have tested enough people"* [P16]<br>*"So if I knew… 'cause I'm in Fife, so I'm Dunfermline. And if, like, say Edinburgh's at 300 just for example, and Fife's at 15 and I work in Edinburgh, I'm not going to travel to Edinburgh because I can specifically know that Edinburgh's really bad out of this."* [P7] |
| Location Specific | Located information made pandemic feel more real – could cause more stress.<br>High local infections strong driver of behaviour change.<br>Could support increased normality | *"If I see this information, I may not choose to go out frequently. It said there is potential risk around maybe my home, my house or other… so I may choose not to go out frequently or play with my friends."* [P17]<br>*"I think if you're just talking about England or Scotland or Wales then you might not necessarily think that it's literally the next road"* [P8]<br>*"But the way the maps look, the idea, the premises behind them, the idea of actually keeping a normality in terms of where you can go and keeping the numbers down, … I really like this for that reason."* [P7] |
| Structured Advertising | Encouragement to resume doing 'normal' things was appreciated.<br>Suggestion to go to a restaurant reassuring to some students but others offended, felt it was out of scope | *"Again, it's important to, you know, support and keep normality to where… the best ability, so I like this app. It's letting me know that… it gives me information that it knows that the areas is, I guess, safe and it's letting me know to go and carry on my life as normal as possible."* [P7] *"I do not want this stuff. This stuff starts to feel like advertising, adverts for restaurants. I understand that we're slightly being pushed to do it, it's an economic incentive, but personally, I choose who I shop with."* [P2] |
| Behavioural Suggestions | Reassurance for resuming normal activities valued<br>Seemed more invasive and controlling the UK | *"It gives some advice about doing an exercise and get some fresh air and I think give me the positive feelings yeah."* [P19]<br>*"I don't want to have advice from this app to, like, do stuff with my life, like the restaurant."* [P3]<br>*"because a lot of the parks near me were crazy busy during peak times, so to me that's a nice idea, it feels safe to go outside and stuff and just a general reassuring feeling I think."* [P5] |
| All notifications | Good information would encourage installing and keeping the app | *"But I think actually things like that map that's on the screen now, they'd sway more people to get the app in the first place, I think, 'cause that's actually a really good feature. And localised information about the R rate, I think, is also good and relevant."* [P4] |

**4.2 Do students support digital contact tracing?**

Most participants felt that having an app of this sort was beneficial, whether because *"This app is really kind"* [P21] , or *"it's a good motivator for basically making people feel that they're doing something"* [P7] . There was a general willingness to install as *"If we don't have a vaccine, then I think a tracing system is the way to go."* [P6] Presenting additional information was a positive thing, *"making sure that people are aware of what's going on and be informed"* [P8] , as was a sense that it was *"safe to go outside and stuff and just a general reassuring feeling I think."* [P5]

Personal protection was a motivator for some, as *"it can protect me from any other persons who are contracted with the virus"* [P18] , however, this particular cohort tended to see the intervention as a way to help others, they *"understand* that I'm not as at risk as everyone else and therefore I have a duty to make sure that I'm not an asymptomatic person who's running about getting other people sick. " [P2]

Table 4. Main findings from showing potential alerts around contact with infected people (Figure 4).

| Function | Main findings | Key quotes |
| --- | --- | --- |
| **Contact Notifications** | | |
| Simple Alerts | Simply stating a contact happened was not convincing. Advice to be more careful appreciated by UK students but not Chinese | "I think once I see this information, I want to know when, where, I want to know more information, If I had a contact, I want to find the reason why I contact him." [P13] "I would consider this message. Because I have a contact doesn't mean I got the coronavirus." [P20] "I think if you don't go into detail, like how do you define what the contact is… So if I was walking past someone in a shop would that be a contact?" [P10] |
| Detailed Alerts | Support needed to calculate how long to self isolate for based on contact date. Detailed timing raised idea of who the contact was, which UK students were dubious about. | "it tells me the date when I have been in contact with potent…, with people and especially with the day, and I could start counting the days from then." [P1] "I can start to work out my list of people who have been sick. But the weird thing is, would I like that sort of information? Yes. But do I actually agree with it ethically? No." [P2] "I can think back about the food shop and think whether I have been careful or not when I was shopping, and then have my own self-assessment, or whether, I think, I don't know, I forgot to wash my hands, I didn't sanitise, that would make me more at risk." [P1] |
| Personalised Alerts | Personal risk factors strongly appreciated by some. However, hard to understand for others. | "I think the fact that it takes into account your personal risk level is useful, because the app will kind of cater its suggestions to how likely you are to get seriously ill or anything. So, I think that's really useful information." [P9] "This might just be my dyslexia playing up but I'm not quite… yeah, I'm not quite understanding what it's trying to tell me." [P7] |

There was a sense that it was virtuous to install the app, whether *"I just feel it's my moral obligation."* [P7] or a sense of *"Oh I'm really happy and I'm proud cos I can have not only for me … but also I help others."* [P11] Quotes from Chinese participants talked more about civic duty and a sense that *"it is like official app, so maybe the government need us to do that"* [P19], while UK participants tended more towards morals and care, as *"having it will protect other people around me as well. … I think it would really just be to look after the people around me."* [P9]

For some, the app was seen as part of a pathway towards a better quality of life, as *"it would allow me more freedom as we lift lockdown"* [P6] and was a *"good way for us to contribute myself to help to reduce this bad situation"* [P17]. This was supported by some of the messaging, and information provided, such as P15 who would be *"glad if I see, we are getting this under control"*. There was a hope that the app could contextualise what was happening, *"to, like, put it as part of the bigger picture"* [P4], as *"it can give me some information about the condition around my neighbourhood, around my city, around my country"* [P16], with the ultimate hope that *"You can observe your local area but also the country as a whole, and then also the UK as a whole, as it slowly transitions into, hopefully, eradicating the disease"* [P2]

### 4.3 What concerns do students have about digital contact tracing?

The installation screens at the start of the application Figure 1 were designed to surface initial concerns that users had; the information provided was *"more more than many apps in China"* [P15] *"felt it was really clear"* [P7] *"I don't think I've ever downloaded an app and seen something so thorough."* [P11]

Bluetooth and battery life were concerns for many participants, such as P19 who was *"not happy for [Bluetooth being enabled]. As I have explained, that I don't like to open my Bluetooth"*. The daily notifications (Figure 2 and 3) prompted discussion about maintaining participation *"as this app will continue tracing it firstly remind me that the condition of Covid 19 still exist, I still need to be careful"* [P21] . This could feel stressful, as reminders that the app was working in the background *" would make me want to keep going back and checking my phone every five minutes."* [P11]

*4.3.1 Privacy issues.* UK 6 thought that *"if I get the indication that this is a surveillance app and policing app then that would make me really uncomfortable, and make me want to want to uninstall it"* . P7 thought that they'd uninstall *"if it started tracking things I was suspicious of"* . When asked about problems with permissions requested as a whole, the most detailed responses were from the UK participants who had multiple issues with *"sharing information commercially and storing it for 20 years"* [P4] and *"giving your information to private companies such as G4S"* [P11] . Some of the more egregious permissions made P7 *":"* [feel like ... it wasn't for its purpose. It made me question actually every single setting] This was often seen as a tradeoff, e.g. *" I completely agree with the principle of the app and I think something definitely has to be in place, but giving all this sort of location information to the government does kind of scare me in a little way."* [P5] The debate about centralised vs. decentralised did not play a prominent role in discussions, although there was a tendency for Chinese participants to *"prefer the centralised model because I think my privacy can be protected"* [P18] or for greater efficiency, while UK participants would *"be more comfortable with decentralised"* [P7] .

*4.3.2 Trust in data Use.* P10 would be unhappy *"if I felt I could be… in trouble for what the NHS had seen on my app"* . He was concerned the app would mean he would *"not have my freedom"* , and that the data might be used for *"issuing a fine"* . There was a key requirement, particularly in the UK to know what data would and would not be used for: *"I probably would have wanted more information just on the things about the data storing as to why different organisations for data to process and also why the other branches of government would need the data. Like with more information I definitely could get onboard but I wouldn't make that decision blindly"* [P8] . This is in contrast to P18: *"the app is developed by the government, so I think we trust the app."*

*4.3.3 Trust in functioning.* Many participants were concerned about how well the Bluetooth tracking captured proximity, for example if *"I'm on the third floor, he's on the fifth floor, but I think we've been same place, something he has touched maybe later I have touched or maybe he coughed at some place and the virus is spread…"* [P13] . There were concerns about how cases were tested and confirmed, with Chinese participants expressing particular worries about whether reported cases were correct: *"At least in China we will have a lot of measurements, first we use a nucleic acid test. We will test the antibody, and some CT results. So if we even to decide one person is real patient we need to combine both results so I think it could be improved result."* [P20] This was highlighted when participants were presented with demands to self isolate: *"Because I have a contact doesn't mean I got the coronavirus. And also to find out, what kind of contact."* [P20]

*4.3.4 Legibility.* While the purpose of the app was generally clear, some of the above statements about keeping Bluetooth off, or using the app only *"at your own discretion"* [P11] implied that a few participants had misconceptions about function. The most divisive screen was Notification 4, (Figure 4) that showed contact level in the context of a personal risk model; this was powerful for several users as *"because it tells me on what, the advice is based"* [P1] , but others were upset that *"it gives me too much information, I need to digest them"* [P18] , or needed other presentations: *"This might just be my dyslexia playing up but I'm not quite… yeah, I'm not quite understanding what it's trying to tell me."* [P7]

*4.3.5 Tone.* There was a sensitivity, particularly in the UK respondents to the tone and nature of the relationship. Users liked notifications that were *"straight to the point"* [P5] , that said *"everything's fine, we're still doing our job"* [P2] and *"being able to know about your area."* [P2] Empty messaging such as 'we are getting this under control' was not appreciated *"because I think that's like a sweeping statement"* [P10] . When the app suggested a safe restaurant to eat at, half of the UK respondents were offended at the overreach, while most of the Chinese respondents were *"glad to receive such news and it gives me specific, like, advice or suggestion that I can go to the restaurant."* [P18]

*4.3.6 Control.* As the app was to be installed on personal devices, questions of control and how it shifts were important to some. Asking participants if the app could switch on Bluetooth automatically provoked reactions, e.g. *" So I want to, if I want to use this app, I will open my Bluetooth, but if I don't want to use it, I just shut it down."* [P15] , or P11 who felt "you put it on at your own discretion opposed to having it run in the background." P13 in particular felt that *"I don't need the app to control me too much, it's just an app"* [P13] . This question extended to the services behind the app, for example *"[the app] tells me to self-isolate, but how do they know that I'm doing that?"* [P14]

## 4.4 Why would students uninstall a contact tracing app?

In contrast to concerns about whether to begin participating in the app, participants also talked about reasons why they might uninstall a personalised digital contact tracing app. Several participants pointed out that they would uninstall the app *"if Covid was finished"* [P2] . However, beyond this there were concerns relating to security, usability, privacy, and trust.

P1 thought she would uninstall the app if data was *"being misused"* or *"if it's too much information being collected then I might delete the app."* P2 also thought he would *"immediately delete [if] there is a data leak or there is some problem that's been noticed about how the app was implemented."* P4 would uninstall *"if something broke in the news about like, a security breach or information being leaked."*

Practically, P3 pointed out that she might temporarily delete the app *""if I need some space ... but then install it later.""* Similarly P8 thought that if covid-19 *"became significantly less of a risk … and I needed to free up some space on my phone."* P9 said something similar about needing to know that the app was not *"just taking up space on my phone"* .

P11 thought she would uninstall the app *"if it didn't feel like it was useful"* or *"if it was starting to malfunction"* . P4 though they might uninstall if they got *"several notifications a day, I think that would just get annoying"* . P4 felt especially likely to uninstall if the app gave *"negative news every day"* . P7 also worried about the app *"telling me every single day about coronavirus."*

## 4.5 Would students comply with instructions to self isolate from a contact tracing app?

When looking at notifications instructing the user to self isolate, attitudes towards compliance varied. The notifications often caused stress: *"I think it would terrify me [...] Because it says that I've had a contact in the last few days, so that would mean that I'm in danger"* [P18] . Many people understood this as a part of their civic or moral duty, and would self isolate *"if it's really necessary, every time it's necessary, yeah"* [P3] , while noting that obeying an app *"if you've done it once it probably would get harder the next time and the next"* [P4] , and eventually be *"more inclined to, kind of, demand* *a test maybe and after three or four [self isolations]"* [P5] . Initially, however, *"the inconvenience of staying in for another two weeks far outweighs any... It's not even a consideration in the fact that I need to stay in just to try and save lives."* [P2]

The Chinese participants were more disposed to question the urgency – *"Do I need to self isolate for now, or just err, tomorrow."* [P11] – the basis of contact notifications – *"I would consider this message. Because I have a contact doesn't mean I got the coronavirus. "* [P20] – or tried to rationalise non-compliance: *"I think for some of people, or just for me, if my lifestyle is really quick, or if I have some deadlines, or dues to finish, I may not listen what an app is saying. So I think maybe you can provide some more detailed information to tell people how serious the situation is"* [P21] .

There was a general demand for clarity of demands – *" I think generally people need to be told do this or don't do this. I think giving people the choice doesn't really work."* [P6] – but also on how they should be carried out: *"Well, maybe it can give you some... how to self-isolate or something like that."* [P18] However, this was coupled with reacting differently to the presentation of the message. Local context supported compliance, for example *"knowing that I either was next to someone who had coronavirus or near someone that had coronavirus, if there was a big spike in a certain area where I was, I would absolutely stay home."* [P7] , with several participants willing to interpret the strength of connection themselves: *"If I did a risky behaviour as to touch the patients, I will choose to isolate myself for two weeks. But if I just go to the shop and buy something and not talk to anyone just someone like near me, I will think its maybe not a big deal"* [P20] . Specific information was seen by many as giving more weight to the contact notifications:

> "So if I'd found out I'd been in close contact with someone, so without the food shop visual or the time shown I don't know if I would really change my behaviour that much because I think I'd just assume that I was in the same shop, like on the opposite end of the shop with someone. You don't want to have the inconvenience in your life. But I think if I saw the specifics it makes it a lot more real. I wouldn't self-isolate, realistically, but I'd be a lot more vigilant." [P10]

There were several other cases where softer behavioural changes might be made, e.g. *"I'd probably not self-isolate if it's not confirmed, I'd maybe stop going to the pub or cafes or things like that, maybe scale back my leisure activities."* [P5] , or a feeling that *"the app would more have to say, go do a test, you know, but be careful, you know, not impose me to just self-isolate"* [P3] . Compliance could be helped if the app could provide detailed information on *"how do I tell my employer that I'm going to be off for two weeks? Am I going to get paid?"* [P7] , or *"a contact number, something like that and further information."* [P11]

## 5 DISCUSSION

We have presented qualitative findings from the interviews. We will now discuss these findings with relation to acceptability factors, and we will then make design recommendations. We do this in order to better understand the "barriers to effective deployment"[3] of digital contact tracing and to inform digital approaches that can help digital contact tracing apps "to achieve wide and consistent use"[3]. Our work focuses on students.

### 5.1 Acceptability of digital contact tracing

To structure discussion of acceptability, we draw on the Theoretical Framework for Acceptability (TFA) [40], which can be used to assess prospective acceptability of digital health interventions [39]. The TFA gives a holistic way to look at some key considerations about the ways that subjects relate to interventions – not simply around ethics, but also ease of use, belief in how well they work and how well they are understood.

*5.1.1 Affective attitude.* The first dimension of the TFA concerns: *"how an individual feels about taking part in an intervention."* There was a general sense of positivity about the idea of contact tracing; most of these participants saw it as a way for them to help others, although some were motivated by personal safety and a desire for protection. There was a combination of altruism and duty as motivations, along with a sense of moral obligation. Supplying extra information, either about what to do, how the response was progressing or what was happening in the local area increased desire to participate. Reassurance was also a strong factor, both for a general sense of safety and guidance about specific activities that were allowed and/or safe.

*5.1.2 Burden.* This concerns: *"the perceived amount of effort that is required to participate in the intervention."* As with most phone based interventions, a key concern was battery life and in particular the perceived effect of enabling Bluetooth. Other technical factors (e.g. storage space) were much less important, with most participants glossing over them, and most saying that they would continually carry their phones when leaving the house. A significant factor of concern was the creation of stress and anxiety, particularly via contact notifications, but also from anything that continually reminds people of the need to be cautious – although this was also seen as promoting safety and useful behavioural modification.

*5.1.3 Opportunity cost.* This concerns: *"the extent to which benefits, profits, or values must be given up to engage in an intervention."* The largest opportunity cost in this sense is the sharing of personal data, with most participants sensitive to what data was to be shared. There were strong boundaries for many participants about who the data should be shared with, and which sharings violated their personal values - in particular, UK respondents were put off by non-governmental organisations processing their data. However, in most cases, it was acknowledged that a level of data sharing was a key component of the intervention, and participants were happy to make that trade.

*5.1.4 Ethicality.* This concerns: *"the extent to which the intervention has good fit with an individual's value system."* The ethics of the base intervention were in line with most respondent's value systems. The main concerns were around the purposes for which personal data might be used. This included loss of personal privacy, but also questions about the purposes for which other branches of the government might use information. Participants were sensitive to the implications of being able to discover who contacts were and the implications of being able to show that you had not had a contact. The other key ethical conflict was between values of autonomy and the way the app functioned – automatically turning on Bluetooth was a key example of overreach, but many users were concerned about the app taking control.

*5.1.5 Self efficacy.* This concerns: *"the participant's confidence that they can perform the behaviour(s) required to participate in the intervention."* The main behaviours asked of the participants were to keep the app installed, carry their phone, and obey injunctions to self isolate. Most were happy to have the app installed, although several made it clear they would turn it off and on as desired. Unsurprisingly, going into self isolation was a sticking point for many. Several users said it would be challenging financially or emotionally, and several others either did not know exactly what was required, or did not have a practical strategy in place for doing it. Extra information was helpful here, with local information helping to make the situation more real, and personalised information helping those who understood it to contextualise their behaviour and support necessary changes.

*5.1.6 Intervention coherence.* This concerns: *"the extent to which the participant understands the intervention, and how the intervention works."* There were some unintended conceptions of the application, such as it might highlight which people around you have Covid-19 and show location traces. Most participants developed a clear understanding of the base purpose quickly, but it is clear that some need a more explicit story about the purpose, and more connection to current practices such as health QR codes or collection of personal details in restaurants. Several participants had the sense that they could just turn the app on when they wanted to, or use it without Bluetooth enabled, which implies they saw it more as a personal information solution than a pervasive surveillance technology.

*5.1.7 Perceived effectiveness.* This concerns: *"the extent to which the intervention is perceived as likely to achieve its purpose."* Most participants had a sense that this kind of intervention was helpful. They had questions about how many others might use it, and how it fitted into the bigger picture of things that were being done. Several participants questioned the functioning of the proximity testing, especially if they were being asked to self isolate. Most students felt that this was a key part of protecting others, which they saw as the primary purpose for them of digital contact tracing.

## 5.2 Design considerations for digital contact tracing

In this study, we have looked at the student population, including 'international' students. There was an overall sense that they supported the intention of contact tracing, but it was less clear how likely they were to comply with instructions from the app to comply. Since this study took place the NHS Covid-19 app for England and Wales has 56% uptake and has issued over 1.7M notifications to self isolate [14]. Less insight is available for engagement with the other UK apps Protect Scotland and StopCovid NI.

By design, the Covid-19 app does not collect data about individual backgrounds and their actions – for example it is not known how many people comply with self isolation instructions, or how stringently, and it is not known how well the app connects with different communities and demographics.

To improve understanding of how digital contact tracing systems are used and to support further development, we offer a collection of design factors that may help to create acceptable digital contact tracing systems that maintain higher engagement with the student population, and a greater chance of advice being followed.

- **Information can support engagement:** The informational screens within the prototype were a reason participants thought they would keep using the app, and was seen as a way to convince people to sign up. Some notification that the app was actually functioning was welcomed by most participants, especially if presented in a supportive manner. A notification that there had not been any contacts was a sign that something was happening, and a reassurance that nothing was amiss. This also works with the user values around civic duty, *"making people feel that they're doing something, doing well"* [P7] . Reasons for uninstalling the app included *"if the situation in my area is not really bad"* [P21] , or *"if the app does not work or does not let me know it works"* [P16] so *"a notification saying it's still running are so important because it then just reminds me that it is actually doing something"* [P9] .
- **Personal information is compelling:** Several participants perceived the effectiveness of notifications that showed an individual risk model would be *"much more than the other ones because of the sort of personalised advice"* [P11] . Showing e.g. a level of contacts so that one could *"know a sort of value in terms of how at risk I am"* [P2] allowed for smaller behavioural changes such as reducing time spent in populated areas, and it was felt that being more transparent about contact and risk levels meant the app would be more convincing when telling people to self-isolate.
- **Geospatial information is powerful:** While the app might work in the background or provide general notifications, geospatial information was a powerful driver mentioned by many respondents as *"providing users with updated information regarding regions or maybe even the local area is probably way more beneficial than high level messages just in text"* [P2] . This was seen as a way to maintain engagement, but also would *"sway more people to get the app in the* *first place"* [P4] and aid word-of-mouth encouragement to install. This can be seen in relation to current offerings in the UK that have begun to integrate geospatial information to a varying degree – the NHS app shows current guidelines through a button on the home screen of the app, while the Protect Scotland app relies on users visiting the website and inputting their postcode.

**Notifications can be stressful:** While some users were reassured by the fact the app was working in the background, constant reminders of the pandemic were seen as both stressful and *"if it was giving me, like, negative news every day as well I'd probably [uninstall the app]"* [P4] . This is distinct from the technical malfunction that cause ghost notifications with the UK tracing app [4], in that notifications are a planned part of a design. There were two aspects to this. Firstly, participants were sensitive to proportionality, and did not want to be disturbed when there were not many cases around, as it would become tiresome, contributing to notification fatigue. Secondly, each notification was a reminder of the pandemic, triggering a sense of negativity, and could be trigger for uninstalling the app. Some, but not all, participants were positive about the use of supportive or encouraging notifications, indicating that end-user preferences for the type and frequency of notifications would be beneficial.

**Design for populations:** there were some extremely divergent opinions about what would be useful or desirable in an app. On a surface level, while detailed information was desirable for some, others preferred graphics, numbers and simple text. To be inclusive of people who find reading English challenging, it is important to have simple presentations backed up by further information. On a deeper level, there were various values reported for complying with the app – civic duty, personal safety, care for family – which would lead to different mechanisms to promote and sustain engagement. This indicates a degree of customisation would be useful, in terms of amount of information presented and style of presentation. Participants diverged strongly on whether the app should give suggestions for e.g. local restaurants to eat at, with some finding it a key feature that would support participation and others an egregious overreach. An option to opt-in to a range of different notifications and notification styles seems necessary to bridge these divergent value systems.

**The app needs to build trust:** The moment of being asked to self isolate is a watershed moment for trust in the system, and highlighted ways in which they did not entirely trust the system. This is a moment when the participants values around altruism and civic duty conflict with the various burdens of having to self isolate, and many participants started to question whether really should follow the guidance. Trust was needed in the necessity of self isolation, to avoid thoughts that *"because I have a contact doesn't mean I got the coronavirus"* [P20] , but also in the technical chain behind the notification – participants wanted contextual evidence that this was a 'real' notification, which related to where and when they were exposed with a sense that the contact was correctly determined. This relates to the app giving indications that it is functioning correctly – for instance, showing daily contacts with non-infections people can build trust that the Bluetooth is not picking up neighbours through the wall.

**Clarity of data use:** Participants' concerns about who might do what with their data took precedence for them over questions about technical mechanisms (e.g. the centralised versus decentralised debate). A sense that something nefarious was being done with their data or any *"indication this is quite a surveillance app and policing app"* [P6] were key reasons to uninstall. There were worries around collection – *"if it started sending out data or working in backgrounds maybe times it didn't need to be working"* [P7] , around whether *"data was being sold on"* [P9] and around the uses to which the data was, whether one could be *"trouble for stuff that the NHS had seen on my app"* [P10] . It is important that contact tracing apps provide clear messaging about the what that data is collected, shared and used.

### 5.3 Limitations of the study

This study was carried out relatively rapidly with an opportunistic, non-representative sample to make information available in a timely manner. We used wireframes rather than developing a functioning application for participants to live with, which also allowed us to explore the design space more fully, although did not provide the sense they were installing on their own device, or experience of living with it. However, from the interview transcripts, users were treating the application as potentially real, and we were able to gather a significant amount of grounded reflection from them. The study was carried out online, which was not a particular barrier, and also made it more natural to include remote participants on an equal footing, so that we could explore different contexts.

## 6 CONCLUSION

Contact tracing apps occupy a difficult design space – they are invisible most of the time, popping up mainly to ask users to do something they do not want to do, while being a more overt form of surveillance than most other applications. In order to maximise public health benefits, it is necessary that a large number of people install, keep and comply with the the app. This paper carries out a design oriented exploration of the factors that go into the end-user acceptability of contact tracing apps, with a view to supporting the development of apps that have good uptake and are effective at informing appropriate behaviours such as self-isolation for current and future pandemics. We have uncovered a range of views and beliefs that can help to make these systems more acceptable, around the presentation of information, the relationship that people build up with the app and the tradeoffs people make between their values of autonomy and privacy and their sense of duty or moral obligation to take part in health surveillance. This has been carried out using an acceptability framework to take into account a wide range of concerns: the values of altruism, duty and personal safety behind joining the effort; the various burdens and costs around participation; ethical stances balancing public good with privacy; and the ways that participants understand and trust the app and the wider contact tracing effort.

By working with students who have experienced severe disruption in their lives, from divergent backgrounds, we find a broad set of perspectives on digital contact tracing, which leads to a set of design guidelines about how to engage publics and create systems that are not just functional, but are *legible, acceptable and humane*.

## ACKNOWLEDGMENTS

This work was funded by the Data Driven Innovation project as part of the Edinburgh City Deal. Many thanks to our anonymous participants, and to all of the Institute for Design Informatics.